\documentclass[12pt]{article}

\newcommand{\be}{\begin{equation}}\newcommand{\ee}{\end{equation}}
\newcommand{\bea}{\begin{eqnarray}}\newcommand{\eea}{\end{eqnarray}}
\newcommand{\nn}{\nonumber}\newcommand{\p}[1]{(\ref{#1})}
 \newcommand{\lb}[1]{\label{#1}}
\newcommand\s{\scriptscriptstyle}

\newcommand\q{\quad}
\newcommand\qq{\qquad}

\newcommand{\Tr}{{\rm Tr}}

\def\cA{{\cal A}}

\def\cV{{\cal V}}
\def\cU{{\cal U}}
\def\cW{{\cal W}}

\newcommand\Ea{\eta^\alpha}
\newcommand{\Bea}{\bar{\eta}^{\da}}

\newcommand{\da}{{\dot{\alpha}}}
\newcommand{\db}{{\dot{\beta}}}

\newcommand{\tta}{\theta_2^\alpha}

\newcommand{\tha}{\theta_3^\alpha}

\newcommand{\btoa}{\bar{\theta}^{1\da}}

\newcommand{\btta}{\bar{\theta}^{2\da}}

\newcommand\adb{{\alpha\db}}
\newcommand\ada{{\alpha\da}}

\newcommand\padb{\partial_\adb}
\newcommand\pada{\partial_\ada}

\newcommand\A{{\s A}}
\newcommand\B{{\s B}}

\newcommand\M{{\s M}}

\newcommand\Y{{\s Y}}

\newcommand{\Dot}{D^1_2}
\newcommand{\Dto}{D^2_1}
\newcommand{\Doh}{D^1_3}

\newcommand{\Dht}{D^3_2}
\newcommand{\Dth}{D^2_3}

\newcommand{\pot}{\partial^1_2}
\newcommand{\pto}{\partial^2_1}
\newcommand{\poh}{\partial^1_3}

\newcommand{\pht}{\partial^3_2}
\newcommand{\pth}{\partial^2_3}

\newcommand{\Vot}{V^1_2}
\newcommand{\Vto}{V^2_1}
\newcommand{\Voh}{V^1_3}

\newcommand{\Vht}{V^3_2}
\newcommand{\Vth}{V^2_3}

\newcommand{\Do}{D^1}
\newcommand{\Doa}{D^1_\alpha}

\newcommand{\Dt}{D^2}
\newcommand{\Dta}{D^2_\alpha}

\newcommand{\bDo}{\bar D_1}

\newcommand{\bDt}{\bar D_2}
\newcommand{\bDta}{\bar{D}_{2\da}}

\newcommand{\bDh}{\bar D_3}
\newcommand{\bDha}{\bar{D}_{3\da}}

\begin{document}
\def\theequation{\arabic{section}.\arabic{equation}}
\begin{center}
\Large\bf
B.M. Zupnik~\footnote{Bogoliubov Laboratory of Theoretical Physics, Joint 
Institute for Nuclear Research, Dubna, 141980 Moscow Region, Russia\\
E-mail: zupnik@thsun1.jinr.ru}\\

  N=4 MULTIPLETS IN N=3 HARMONIC SUPERSPACE\\
\vspace{0.5cm}
\end{center}

\begin{abstract}
It is shown that the $N{=}3$ harmonic-superfield equations of motion are 
invariant with respect to the 4-th supersymmetry. The SU(3) harmonics are 
also used to analyze a more flexible form of superfield constraints for 
the Abelian $N{=}4$ vector multiplet and its $N{=}3$ decomposition. An 
alternative unusual representation of the  $N{=}4$ supersymmetry is 
realized on infinite multiplets of analytic superfields in the $N{=}3$ 
harmonic superspace. U(1) charges of superfields in these multiplets are 
parametrized by an integer-valued parameter which plays the role of the 
discrete coordinate. Each superfield term of the $N{=}3$ Yang-Mills action 
has the infinite-dimensional $N{=}4$ generalization. The gauge group of 
this model contains an infinite number of superfield parameters. 
\end{abstract}

\setcounter{equation}0
\section{Introduction}
The Yang-Mills theory in the harmonic $D{=}4,~N{=}3$ superspace was  
constructed in Ref. \cite{GIKOS3} (see also  \cite{GIO,DM} and book 
\cite{HS}). In this approach, physical component fields of the $N{=}3$ 
vector multiplet are included into geometric superfields which have an 
infinite number of auxiliary fields off mass shell. The superfield 
geometry of the $N{=}3$ gauge theory is based on the preservation of the 
Grassmann analyticity, and superfield equations are zero-curvature 
conditions for basic superfields (harmonic potentials) of this theory. 
The off-shell structure of $N{=}3$ harmonic superfields plays an important 
role, in particular, for constructing the $N{=}3$ analogue of the 
Born-Infeld action \cite{IZ}.
  
The $N{=}4$ vector multiplet was  considered in the framework of the
SU(4)-harmonic superspace  \cite{IKNO}-\cite{Zu4}. In this quite 
complicated harmonic formalism, one can construct a harmonic projection 
of superfield strengths satisfying simple equations of motion; however, 
nobody knows how to solve problems of constructing the corresponding 
superfield action. It is not excluded that these problems are connected 
with a manifest SU(4)-invariance and a high dimension of the Grassmann 
part of the $N{=}4$ superspace. We consider a more flexible $N{=}3$ 
superspace and analyze different realizations of the $N{=}4$ supersymmetry 
on $N{=}3$ superfields and equations of motion. In particular, the $N{=}4$ 
generalization of the $N{=}3$ gauge action for an infinite multiplet of 
superfields is proposed. 
 
It is well known that  physical field-components of the  $N{=}3$ and 
$N{=}4$ vector multiplets are completely identical; however, 
transformations of the $N{=}4$ supersymmetry for these component fields 
form a closed algebra on the equations of motion. The $N{=}4$ Yang-Mills 
theory can also be constructed in the $N{=}2$ harmonic superspace 
\cite{HS}, but  the algebra of additional supersymmetry transformations of 
this formulation is closed on-shell only.

Section 2 is devoted to the $N=4$ supersymmetry realized on superfield 
solutions of the $N{=}3$ equations of motion. We construct additional 
spinor transformations of the $N{=}3$ superfield strengths and potentials 
on mass shell. It is shown that the Abelian and non-Abelian gauge 
equations in $N{=}3$ superspace are invariant with respect to the 4-th 
supersymmetry.

Superfield $N{=}4$ equations (constraints) in the standard superspace
can be decomposed in terms of the 4-th spinor coordinate to study the
$N=4$ transformations of the $N{=}3$ gauge superfields and solutions of 
constraints with  hidden SU(4) symmetry. In section 3, we  analyze the 
SU(3) covariant representation of the  $N{=}4$ superfield constraints for 
the Abelian gauge multiplet. One can try to use the SU(3)/U(1)$\times$U(1) 
harmonics in order to understand the geometric meaning of these equations
and their relations with the $N{=}3$ harmonic superspace. Decomposition
of $N{=}4$ superfields in the 4-th spinor coordinate allows us to study
$N{=}4$ transformations of the gauge $N{=}3$ superfields. It should be 
said that we do not find an appropriate action without Lagrange 
multipliers in this approach.

In Ref. \cite{BIK}, the $N{=}4$ supersymmetry with central charges was  
realized on an infinite set of chiral $N{=}2$ superfields which then could  
be nonlinearly expressed via the single chiral $N{=}2$ superfield 
describing the partial spontaneous breaking of the $N{=}4$ supersymmetry. 
In the SU(4) harmonic formalism \cite{Zu4}, one can also define geometric
$N{=}4$ superfields whose decompositions contain infinite multiplets of 
$N{=}3$ superfields of different dimensions. The $N{=}3$ harmonic
superspace gives us a  simpler possibility to realize transformations of 
the $N{=}4$ supersymmetry and the corresponding version of superfield 
geometry. In  section 4, we study the $N{=}4$ transformations on an 
infinite set of $N{=}3$ superfields parametrized by an integer-valued 
parameter $k$ which plays the role of the auxiliary discrete coordinate.
The 4-th spinor transformation of this set contains the spinor derivatives
which preserve the Grassmann analyticity. The multiplication rule for 
these infinite $N{=}4$ multiplets is used  to construct an $N{=}4$ 
invariant generalization of the  $N{=}3$ Yang-Mills action. We analyze 
also the gauge invariance of this action. The classical equations for  
additional $N{=}3$ superfields may have nontrivial solutions, so the 
consistency of the infinite-superfield model at the quantum level and 
its relation with the $N{=}3$ gauge theory require an additional analysis.
A possibility of 'truncated' realizations of the $N{=}4$ supersymmetry
on a finite number of $N{=}3$ superfields is also considered; however,
superfield actions for these realizations are not constructed.  

In the appendix, we give a review of basic formulae of the $N{=}3$ gauge 
theory in the harmonic superspace which are used in other sections of the 
paper. 

\setcounter{equation}0
\section{ N=4 transformations of N=3 superfield equations }

Supersymmetry transformations of the components of the $N{=}4$ vector 
multiplet and their equations of motion are well known. In this approach, 
the algebra of all supersymmetry transformations is closed on the 
equations of motion. Formulations of the $N{=}4$ Yang-Mills theory in 
terms of $N{=}1$ or $N{=}2$ superfields correspond to a manifest 
realization of the corresponding supersymmetries.

We shall use the formalism of the $N{=}3$ gauge theory in the harmonic 
superspace, which is reviewed in the  appendix. In the $N{=}3$ superspace,
three supersymmetries are realized manifestly on the gauge superfield 
potentials  $V^I_K$ \p{lambda} and superfield strengths $W^1,~\bar W_3$ 
\p{nonal}, so we shall study a possible connection with the 4-th 
supersymmetry. Let us consider solutions of the harmonic superfield 
equations of motion $\Vot|~,~\Vth|~,~W^1|$ and  $\bar W_3|$ and the 
corresponding transformations of additional supersymmetry. Superfield 
solutions of free equations will be marked by symbol $|_0$. One can define 
the following  spinor transformations on the free superfield strengths of 
the Abelian $N{=}3$ theory:
\be
\hat\delta_\eta W^1|_0=-\Bea\bDta \bar W_3|_0~,\q
\hat\delta_\eta\bar W_3|_0=\Ea\Dta  W^1|_0~.\lb{Wonsh}
\ee
The free equations  \cite{AFSZ}
\be
(\Dot, \Dth)(W^1 , \bar W_3)=0~,\lb{harman3}
\ee
 are invariant with respect to these  spinor 
transformations
\be
\hat\delta_\eta(\Dot, \Doh) W^1=(\Dot, \Doh)\hat\delta_\eta W^1|_0
=-\Bea\bDta(\Dot, \Doh) \bar W_3|_0=0~.
\ee
By definition, transformations \p{Wonsh} satisfy the following conditions 
characterizing the free superfields: 
\be
(\Dt)^2\hat\delta_\eta W^1|_0=0~,\q (\bDt)^2\hat\delta_\eta\bar W_3|_0=0 
~.
\ee 
Additional spinor transformations together with the $N{=}3~$ 
$\epsilon$-transformations form the standard $N{=}4$ superalgebra without 
central charges 
\be
[\hat\delta_{\eta_2},\hat\delta_{\eta_1}]W^1|_0=2i(\eta^\alpha_1\Bea_2-
\Ea_2\Bea_1)
\pada W^1|_0~,\q [\delta_\epsilon,\hat\delta_\eta]W^1|_0=0~.\lb{Walg}
\ee

A spinor transformation of the potential corresponding to the 
transformation $\hat\delta_\eta\bar W_3$ \p{Wonsh} has the following form:
\be
\hat\delta_\eta\Vot|_0=2\eta_\alpha\tta  W^1|_0~.\lb{Vonsh}
\ee
The spinor  transformation of the  potential $\Vth|_0$ can be obtained by
conjugation
\be
\hat\delta_\eta\Vth|_0=-2\bar\eta_\da\btta \bar W_3|_0~,
\ee

In comparison with \p{Walg}, the Lie bracket of the potential 
transformations contains gauge-dependent unusual terms. More general 
transformation of $\Vot$ can contain additional  terms which do not 
contribute to the transformation of $\bar W_3|_0$ \p{Wonsh}, but change 
the algebra of transformations on $\Vot|_0$. 

Spinor transformations of the non-Abelian potentials  can be chosen in the 
following form:
\bea
&&\hat\delta_\eta\Vot|=2\eta_\alpha\tta W^1|(\Vth)~,\nn\\
&&\hat\delta_\eta\Voh|=2\eta_\alpha\tha W^1|(\Vth)-2\bar\eta_\da\btoa 
\bar W_3|(\Vot).\lb{4naV}
\eea
The non-Abelian equations of motion are invariant with respect to these
transformations
\be
\hat\delta_\eta [\nabla^1_3, \nabla^1_2]=2\eta_\alpha\tta\nabla^1_3 W^1|-
2\eta_\alpha\tha\nabla^1_2 W^1|+2\bar\eta_\da\btoa\nabla^1_2\bar W_3|=0~.
\ee
In this calculation, we have used conditions of the covariant harmonic 
analyticity of the superfield strengths \p{nahan}.

It is convenient to study nonlinear transformations \p{4naV} using gauge 
conditions \p{gcond}  and nonlinear expansions of superfields $W^1(\Vth)$ 
and $\bar W_3(\Vot)$ \p{naW} in terms of non-Abelian potentials. In 
particular, these formulae allow us to construct spinor transformations of 
non-Abelian superfield strengths via transformations of the potentials 
\p{4naV} .  

\setcounter{equation}0
\section{SU(3) decomposition of N=4 superfield constraints}

SU(4) invariant harmonic-superspace formulations of the equations of 
motion for the $D{=}4,~ N{=}4$ vector multiplet were analyzed in Refs. 
\cite{IKNO}-\cite{Zu4}. It is not difficult to embed the physical fields 
of this multiplet into infinite-dimensional $N{=}4$ multiplets in the 
harmonic superspace; however, it is not clear how to build the superfield 
action in this formalism.

Let us suppose that the $N{=}3$ theory describes some phase of the $N{=}4$
theory which has no manifest SU(4) symmetry. In order to study relations 
of the $N{=}4$ and $N{=}3$ theories, we consider the SU(3) decomposition 
of a free $N{=}4$ superfield strength $W^{\A\B}$ 
\bea
&&W^{\A\B}=(\bar W^{ik}, W^{i4})~,\q W_{\A\B}=( W_{ik}, \bar W_{i4})
\eea
satisfying the following version of the $N=4$ reality condition:
\bea
&&\bar W^{ik}=\varepsilon^{ikl}\bar W_{l4}~,\q W_{ik}=\varepsilon_{ikl} 
W^{l4}
\eea
where $A=(i, 4)$, $i, k, l=1,2,3$ are the indices of SU(4) or SU(3) 
groups, respectively. In the formalism with  hidden SU(4) symmetry, one 
can treat $W^{i4}$ and $\bar W_{i4}$ as independent $N{=}4$ superfields. 
The superfield equations of the $N{=}4$ Abelian vector multiplet have the 
following form in this setting:
\bea
&&D^i_\alpha W^{j4}+D^j_\alpha W^{i4}=0~,\q D^4_\alpha W^{i4}=0~,\\
&&\bar D_{i\da}W^{j4}-{1\over3}\delta^j_i\bar D_{k\da}W^{k4}=0~,\\
&&2\bar D_{4\da}W^{i4}=\varepsilon^{ikl}\bar D_{k\da}\bar W_{l4}~.
\eea

One can try to study these constraints using the SU(3)/U(1)$\times$U(1)
harmonics and the following notation for conjugated harmonic projections
of the $N{=}4$ superfields:
\be
\cW^1\equiv u^1_kW^{k4}~,\q \bar\cW_3=u_3^k\bar W_{k4}~.
\ee
SU(3)-harmonic projections of the $N{=}4$ constraints are equivalent to
the relations
\bea
&&(D^1_\alpha,  D^4_\alpha, \bDta, \bDha)\cW^1=0~,\q\Dth \cW^1=0\nn\\
&&(D^1_\alpha, \Dta, \bar D_{4\da},  \bDha)\bar\cW_3=0~,\q \Dot\bar\cW_3=0
\lb{twanal}\\
&&\bar D_{4\da}\cW^1=\bDta \bar\cW_3~,   \q D^4_\alpha\bar\cW_3=-\Dta\cW^1
\lb{onsh}\\
&&(\Dot,  \Doh)\cW^1=0~,\q (\Dth, \Doh)\bar\cW_3=0\lb{harman}
\eea 
Equations \p{twanal} can be formally solved off-shell via some $N{=}4$ 
superfields (potentials) by analogy with the solutions of the $N{=}3$ gauge 
constraints \p{nonal}; however, we do not know a geometric interpretation 
of these $N{=}4$ potentials. Relations \p{onsh} give directly the 
equations of motion
\be
(\bar D_4)^2\cW^1=0~,\q (\bDt)^2\bar\cW_3=0~.
\ee
Note that similar  equations
\be
[(D^3)^2, (\bDo)^2, (\bDt)^2]\bar\cW_3=0
\ee
follow also from the harmonic equations \p{harman} and the Grassmann 
analyticity of $\bar\cW_3$ \p{twanal}.

Let us consider a  decomposition of the superfield $\cW^1$ in terms of the
4-th spinor coordinate
\bea
&&\cW^1=W^1-{i\over2}\theta^\alpha\bar\theta^\da\pada W^1-{1\over32}
(\theta)^2(\bar\theta)^2\partial^\ada\pada W^1\nn\\
&&+\bar\theta^\da S^1_\da-{i\over4}\theta^\alpha(\bar\theta)^2\pada 
S^{1\da}+(\bar\theta)^2P^1
\eea
where $\theta\equiv \theta_4$, $\bar\theta\equiv\bar\theta^4$ and $W^1, 
S^1_\da$ and $P^1$ are the corresponding $N=3$ components
\be
(\Doa, \bDta, \bDha, \Dth)(W^1, S^1_\da, P^1)=0~.\lb{n3com}
\ee
 The 4-th
supersymmetry transformations of these $N=3$ superfields are
\bea
&&\delta_\eta W^1=-\Bea S^1_\da~,\q \delta_\eta S^1_\da=-2i\Ea\pada W^1
+2\bar\eta_\da P^1\nn\\
&&\delta_\eta P^1=i\Ea\pada S^{1\da}~.\lb{n3decom}
\eea

The on-shell constraints \p{onsh} yield the relations
\be
S^1_\da=\bDta \bar W_3~,\q P^1=0~,\lb{4onshell}
\ee
which give us the  equations
\be
(\bDt)^2\bar W_3=0~,\q \Dth\bDta \bar W_3=0~.\lb{freeW}
\ee

Equations \p{onsh} can be obtained from a superfield action with 
Lagrange multipliers; however, these superfield multipliers contain 
additional degrees of freedom and do not have a geometric interpretation.
It is also unclear whether one can construct an $N{=}3$ superfield action 
without Lagrange multipliers for the $N{=}4$ supermultiplet $W^1,~S^1_\da$ 
and $P^1$ \p{n3com}.

\setcounter{equation}0
\section{Infinite-dimensional multiplet of N=3 superfields}
\subsection{Discrete coordinate and alternative realization of the 
4-th supersymmetry }

In this section, we define the unusual realization of the $N{=}4$ 
supersymmetry on an infinite set of analytic $N=3$ superfields. Introduce 
first a special notation 
\be
A[k]=(A[-n]~,~A[0]~,~A[n])
\ee
for infinite-dimensional multiplets of $N{=}3$ superfields having
U(1)$\times$U(1) charges corresponding to charges of products of the 
harmonics $u_2^i$ or $u^2_i$
\bea
&&h\, A[k]=-k~,\qq \tilde h\, A[k]=k~,\lb{charges}\\
&&A[-n]\sim (u_2^{i_1}\ldots u_2^{i_n})~,\q A[n]\sim (u^2_{i_1}\ldots 
u^2_{i_n}).\nn
\eea
Note that one can use an alternative notation for these harmonic 
multiindices
\be
A[-2]\equiv A_{22}~,\q A[-1]\equiv
A_2~,\q A[0]\equiv A~,\q A[1]\equiv A^2~,\q A[3]\equiv A^{222}~.
\ee 

Transformations of the 4-th supersymmetry can be realized on three 
infinite-dimensional multiplets of analytic $N=3$ superfields:
\bea
&&V^1_2[k]\equiv(\ldots V^1_{222}~, V^1_{22}~, V^1_2~, V^1~,~ V^{12}~,~ 
V^{122} ~,~V^{1222} \ldots)\lb{aset}\\
&&V_3^2[k]\equiv(\ldots V_{2223}~,~ V_{223}~,~  V_{23}~,~ V_3~, V_3^2~,~
   V^{22}_3~,~  V^{222}_3\ldots)\lb{bset}\\
&&V^1_3[k]\equiv(\ldots V^1_{2223}~,~ V^1_{223}~,~ V^1_{23}~,~ V^1_3~,~ 
 V^{12}_3~,~ V^{122}_3~,~ V^{1222}_3\ldots)\lb{rset}
\eea
parametrized by an integer $k$ or a natural number $n$. 'External' charge 
indices correspond to the charges of the $N{=}3$ gauge potentials which 
are included to these multiplets. Let us introduce a common notation for 
these multiplets $\cV[k]\equiv(V^1_2~,~ V^2_3~,~ V^1_3)[k]$ and treat the 
parameter $k$ as the additional discrete (charge) coordinate. 

Each superfield in this set has  definite values of U(1)-charges which 
are defined by the parameter $k$ and the 'external' indices
\bea
&&h\, V^1_2[k]=2-k~,\q\tilde h\,V^1_2[k]=k-1\nn\\
&&h\, V_3^2[k]=-k-1~,\q\tilde h\,V_3^2[k]=k+1\nn\\
&&h V^1_3[k]=1-k~,\q\tilde h\,V^1_3[k]=k+1~.\nn
\eea

Define the following additional supersymmetry transformations for each of 
these multiplets:
\bea
&&\delta_\eta \cV[k]=\Ea\Dta \cV[k-1]-\Bea\bDta \cV[k+1]~,
\lb{4thsusy}
\eea
where $\eta^\alpha$ and $\bar\eta^\da$ are the spinor parameters. The 
given representations contain superfields of the same dimension in 
distinction with the $N{=}3$ decompositions of $N{=}4$ superfields 
\p{n3decom}.

The additional transformations commute with the spinor transformations
of the $N{=}3$ supersymmetry and preserve all properties of the $N{=}3$ 
analytic superfields 
\be
[\delta(\eta,\bar\eta),\delta(\epsilon_i,\bar\epsilon^i)]\cV[k]=0~.
\ee

The Lie bracket of two $\eta$-transformations gives the 4D translations
of all superfields
\be
[\delta_{\eta_2},\delta_{\eta_1}]\cV[k]=2i(\eta^\alpha_1\Bea_2-\Ea_2
\Bea_1)\pada \cV[k]~.\lb{n4alg}
\ee

It is easy to understand that multiplication rules for different $N{=}4$ 
supermultiplets $\cV[k]$ and $\cU[k]$ should include summarizing over the  
parameter $k$, e.g.
\bea
&&\cA[p]\equiv \sum\limits_k\cV[k+p]\cU[-k]~,\lb{product}\\
&&h\, \cV[k]=v-k~,\q h\,\cU[k]=u-k~,\q h\,\cA[p]=v+u-p \nn
\eea
where the integer-number parameter $p$ characterizes the new $N{=}4$ 
multiplet $\cA[p]$, and  $v$ and $u$ are some external charges. One can 
straightforwardly check that the $\eta$-variation of this product has the 
standard $N{=}4$ form \p{4thsusy}
\bea 
&&\delta_\eta \cA[p]=\Ea\Dta \cA[p-1]-\Bea\bDta \cA[p+1] ~.
\lb{Ltr}
\eea
We have used an invariance of infinite sums with respect to translations 
of the  parameter $k$, for instance,
\bea
& \sum\limits_k\{\Ea\Dta\cV[k+p-1]\cU[-k]+\cV[k+p]\Ea\Dta\cU[-1-k]\}=
\Ea\Dta\sum\limits_k\{\cV[k+p-1]\cU[-k]\}~,&\nn
\eea
where the change  $k\rightarrow k-1$ has been made in the 2-nd sum.

\subsection{Truncated representations of N=4 supersymmetry}

Let us consider the transformation of the $N{=}3$ potential in the
infinite multiplet \p{4thsusy}
\be
\delta_\eta \Vot=\Ea\Dta V^1_{22}-\Bea\bDta V^1~.
\ee
We do not know how this representation is related to the realization of 
the $N{=}4$ supersymmetry on mass shell \p{Vonsh}, so one discusses a 
formal possibility of 'truncation' of infinite-dimensional representations
of the $N{=}4$ supersymmetry, when higher terms of supermultiplets
$\cV[k]$ are expressed via a finite number of independent superfields
provided the structure of transformations \p{4thsusy} is preserved.
As the simplest example one shall consider the transformations
\bea
&&\delta_\eta\Vot=c_2(u)\Ea\Dta\Vot+\tilde{c}^2\Bea\bDta\Vot~,\nn\\
&&\delta_\eta\Vth=c_2(u)\Ea\Dta\Vth+\tilde{c}^2\Bea\bDta\Vth~,\lb{trunk}
\eea
where $c_2(u)$ and $\tilde{c}^2(u)$ are formal harmonic series
satisfying a simple relation
\be
 c_2(u)\tilde{c}^2(u)=-c_2(u)\widetilde{c_2}=-1~.\lb{formal}
\ee 
This relation yields an infinite system of equations for coefficients
of the harmonic series. We do not require any regularity for the function
$c_2(u)$, in particular, it cannot be a harmonic polynomial.

The Lie bracket of transformations \p{trunk} has a standard form analogous
to relation \p{n4alg}. Note that the introduction of the functions 
$c_2(u)$ and $\tilde{c}^2(u)$ breaks the SU(3) covariance. These harmonic 
functions allow us to define the truncated solution for the 
infinite-dimensional $N{=}4$ multiplets
\bea
&&(\Vot~,~\Vth)[k]=(c_2)^k(\Vot~,~\Vth)~,\q (c_2)^{-n}\equiv-(-1)^n
(\tilde{c}^2)^n ~.\lb{trsol} 
\eea
We plan to analyze this and other possible truncated realizations
of the $N{=}4$ supersymmetry in further investigations and now
shall continue to study the infinite-dimensional $N{=}4$ multiplets.

\subsection{N=4 generalization of free N=3 superfield action }

A simple geometric interpretation of infinite-dimensional $N{=}4$
multiplets $V^I_K[k]$ as $N{=}3$ potentials depending on the additional
discrete coordinate allows us to construct a classical superfield action
of this model. The superfield action of the $N=3$ gauge theory is 
considered in the appendix \p{n3ym}. The free action of the $N{=}3$ 
Abelian superfield is
\be
S_f(N{=}3)=-{1\over4}\int du\, d\zeta(^{33}_{11}) \{V^1_2D^2_3V^1_3
+V^1_3D^1_2V^2_3+V^2_3D^1_3V^1_2-{1\over2}(V^1_3)^2\}~.\lb{n3free}
\ee
Each term of this action admits the infinite-dimensional $N{=}4$ invariant 
generalization constructed in accordance with the multiplication rule for 
$N{=}4$ multiplets \p{product}. Using the relations
$$
[(\Dot,~\Dth,~\Doh),(\Dta,~\bDta)]\cV[k]=0
$$ 
one can readily show that the harmonic derivatives $(\Dot,~\Dth,~\Doh)
\cV[k]$ are transformed as $N{=}4$ multiplets with modified values of 
external charges \p{4thsusy}. Let us consider the following $N{=}4$ 
multiplets corresponding to all superfield terms of the free $N{=}3$ 
action:  
\bea
&&A^{11}_{33}\equiv \sum\limits_k V^1_2[k]\Dth V^1_3[-k]
=V^1_2\Dth V^1_3+V^1_{22}\Dth V^{12}_3+V^1\Dth V^1_{32}
+\ldots\lb{Aterm}\\
&&\tilde{A}^{11}_{33}\equiv \sum\limits_k 
V^1_3[k]\Dot V_3^2[-k]=V^1_3\Dot V^2_3+V^1_{32}\Dot V^{22}_3
+V^{12}_3\Dot V_3+\ldots\lb{tAterm}\\
&&B^{11}_{33}\equiv \sum\limits_k 
V_3^2[k]\Doh V^1_2[-k]=V_3^2\Doh V^1_2+V_3\Doh V^1+V_3^{22}\Doh V^1_{22}
+\ldots\lb{Bterm}\\
&&C^{11}_{33}\equiv -{1\over2}\sum\limits_k
V^1_3[k]V^1_3[-k]=-{1\over2}(V^1_3)^2-V^{12}_3V^1_{32}-V^{122}_3V^1_{322}
-\ldots\lb{Cterm}
\eea 
All these superfields  are constructed by  analogy with the component 
$\cA[0]$ in the product of $N{=}4$ multiplets \p{product}, so their 
$\eta$-variations are combinations of total spinor derivatives, for 
instance,
\bea 
&&\delta_\eta A^{11}_{33}=\Ea\Dta A^{11}_{233}-\Bea\bDta A^{112}_{33}~,
\lb{Atr}
\eea
The $N{=}3$ superspace integrals of these superfields are invariant with 
respect to the $N{=}4$ supersymmetry.

The action of the $N{=}4$ Abelian multiplet $\cV[k]$ can be constructed as 
an infinite sum of the Chern-Simons-type  superfield terms 
\be
S_f(N{=}4)=-{1\over4}\int du\, d\zeta(^{33}_{11}) \{ A^{11}_{33}+
\tilde A^{11}_{33}+ B^{11}_{33}+ C^{11}_{33}\}=\sum\limits_kS_k~,
\lb{n4free}
\ee
where $S_0=S_f(N{=}3)$  \p{n3free}. This action is invariant with respect 
to the following Abelian gauge transformations:
\bea
&&\delta_\lambda V^I_J[k]=-D^I_J 
\lambda[k]~,\qq J < J~,
\eea
where the analytic gauge parameters $\lambda[k]$  form an infinite 
supermultiplet  which includes  the $N=3$ gauge parameter $\lambda[0]
\equiv\lambda$ \p{lambda}. One can readily check this invariance using the 
relations
\bea
&&\delta_\lambda A^{11}_{33}=\sum\limits_k \lambda[k]\{
\Dot\Dth V^1_3[-k]-\Doh\Dth V^1_2[-k]\}+\mbox{div}\nn\\
&&\delta_\lambda \tilde A^{11}_{33}=\sum\limits_k \lambda[k]\{
\Doh\Dot V_3^2[-k]-\Dth\Dot V^1_3[-k]\}+\mbox{div}\nn\\
&&\delta_\lambda B^{11}_{33}=\sum\limits_k \lambda[k]\{
\Dth\Doh V^1_2[-k]-\Dot\Doh V_3^2[-k]\}+\mbox{div}\nn\\
&&\delta_\lambda C^{11}_{33}=-\sum\limits_k \lambda[k]\Doh V^1_3[-k]
+\mbox{div}\nn
\eea
where the terms with the total harmonic derivatives are not written down 
explicitly.

The infinite set of the zero-curvature equations for the free $N=4$
action \p{n4free} contains three independent series
\bea
&&V^1_3[k]=\Dot V_3^2[k]-\Dth V^1_2[k]~,\nn\\
&&F^{11}_{32}[k]\equiv\Doh V^1_2[k]-\Dot V^1_3[k]=0~,\lb{bas2}\\
&&F^{21}_{33}[k]\equiv\Dth V^1_3[k]-\Doh V_3^2[k]=0~.\nn
\eea
The 1-st  equation is pure algebraic and can be solved off-shell. This 
solution $\hat{V}^1_3[k]$ can be used to define the 2-nd order free 
$N{=}4$ action on the restricted set of independent potentials
\be
\hat{S}_f(N{=}4)=-{1\over4}\int du\, d\zeta(^{33}_{11}) \{ B^{11}_{33}- 
\hat{C}^{11}_{33}\}=\sum\limits_k\hat{S}_k~,\lb{n4sec}
\ee
where $B^{11}_{33}$ is given by Eq.\p{Bterm} and the second term can be 
obtained from \p{Cterm} via the substitution
\be
V^1_3[k]~\Rightarrow~\hat{V}^1_3[k]=\Dot V_3^2[k]-\Dth V^1_2[k]~.
\ee

The structure of solutions to the infinite set of equations can be 
analyzed in non-supersymmetric gauges. Using the component decomposition, 
it is not difficult to show that some branches of the infinite sets 
\p{aset} and \p{bset} are pure gauge superfields
\bea
&& V^1_2[-n]=-\Dot\lambda[-n]~,\qq V_3^2[n]=-\Dth\lambda[n]~.
\lb{lamgauge}
\eea
In this gauge with broken 4-th supersymmetry, the following independent 
superfields remain:
\bea
&&  V^2_3[-n]~,~ V^2_3~,~V^1_2~,~V^1_2[n]~,\lb{n4gauge}
\eea
as well as a 'residual' gauge invariance with the restricted gauge 
parameters
$$
\Dot\hat\lambda[-n]=0~,\qq \Dth\hat\lambda[n]=0~.
$$

The superfield $N{=}3$ equation 
\be
F^{11}_{32}=\Doh V^1_2-\Dot(\Dot\Vth-\Dth\Vot)=0
\ee 
gives in components free equations of the physical fields of the $N{=}3$
vector multiplet  \cite{GIKOS3}
\be 
\phi^k(x), \bar\phi_k(x), A_m(x),  \psi_k^\alpha(x), 
\bar\psi^{k\da},  \psi^\alpha, \bar\psi^\da~.\lb{n3on}
\ee
Auxiliary component fields of the free superfield $V^1_2$ vanish on mass 
shell.

Let us consider  the following pairs of harmonic equations: 
\be
F^{11}_{32}[n]=(\Doh+\Dot\Dth)V^1_2[n]=0~,\q F^{12}_{33}[n]=
-(\Dth)^2V^1_2[n]=0~.
\ee
These equations contain the common superfield $V^1_2[n]=(V^1,~V^{12},
\ldots)$ in  gauge \p{n4gauge} and have  nontrivial solutions for any $n$ 
describing 'short' superfields with the components of exotic dimensions 
and SU(3) representations. For instance, the superfield $V^{12}$ has a 
dimensionless scalar component $u_3^kR_k$. 

\subsection{N=4 invariant  non-Abelian interaction}

We shall construct the $N{=}4$ generalization of the $N{=}3$ interaction
 \p{n3ym} which contains the superfield
\be
J^{11}_{33}= V^1_3[V^1_2,V^2_3]~.\lb{n3int}
\ee
The evident $N{=}4$ generalization of this term  is the following
infinite double sum of superfields
\be
I^{11}_{33}=\sum\limits_{k,p}V^1_3[k]\left[V^1_2[p-k],V_3^2[-p]\right]
\lb{n4int}
\ee
where $p, k$ are arbitrary integers and all superfield potentials
are considered in the adjoint representation of the Lie algebra of
the gauge group $G$.

The $N{=}4$ transformations of this superfield can be analyzed by analogy
with \p{Ltr}, \p{Atr}. Thus, the $N{=}4$ invariant generalization of the 
$N{=}3$ Yang-Mills theory action has the following form:
\be
S_{\Y\M}(N{=}4)=-{1\over4}\int du\, d\zeta(^{33}_{11})\Tr \{A^{11}_{33}+
\tilde{A}^{11}_{33}+B^{11}_{33}+C^{11}_{33}+I^{11}_{33}\}~,\lb{n4ym}
\ee
where the free terms are given by \p{Aterm}-\p{Cterm}. This 
Chern-Simons-type action is invariant with respect to the following
non-Abelian gauge transformations:
\bea
&&\delta_\lambda V^1_2[k]=-\Dot\lambda[k]-\sum\limits_p\left[V^1_2[k+p],
\lambda[-p]\right]\nn~,\\
&&\delta_\lambda V_3^2[k]=-\Dth \lambda[k]-\sum\limits_p\left[V_3^2[k+p],
\lambda[-p]\right]~,\\
&&\delta_\lambda V^1_3[k]=-\Doh \lambda[k]-\sum\limits_p\left[V^1_3[k+p],
\lambda[-p]\right]~,\nn
\eea
where $\lambda[k]$ is the $N{=}4$ set of analytic parameters. The commutator
terms in these formulae preserve the $N{=}4$ structure in accordance
with \p{product}. The proof of invariance is completely analogous to
the corresponding proof in the $N{=}3$ theory \cite{HS} if one uses the
simple properties of infinite sums. 

The equations of motion for the non-Abelian $N{=}4$ action are  the 
infinite-dimensional generalization of the $N{=}3$ zero-curvature 
conditions \p{nazc}
\bea
&&V^1_3[k]=\Dot V_3^2[k]-\Dth V^1_2[k]+\sum\limits_p\left[V^1_2[k+p],
V_3^2[-p]\right]~,\nn\\
&&F^{11}_{32}[k]\equiv\Doh V^1_2[k]-\Dot V^1_3[k]
+\sum\limits_p\left[V^1_3[k+p],
V^1_2[-p]\right]=0~.\lb{nabas2}
\eea

The $N{=}3$ potentials $\Vot,~\Vth$ and $V^1_3$ interact with all 
additional superfields in the infinite  $N{=}4$ multiplets. The problem 
is how to estimate the role of additional superfields in the $N{=}4$ 
invariant phase of this model. 

\setcounter{equation}0
\section{Conclusions}

It is shown that the $N=3$ harmonic-superspace equations of motion are 
invariant with respect to the additional spinor transformation of the 
$N{=}4$ supersymmetry. The corresponding transformations of the 
$N{=}3$ potentials on mass shell have an unusual supersymmetry algebra 
which contains gauge-dependent terms. 

The problem of a manifestly supersymmetric off-shell description of the
$N{=}4$ Yang-Mills theory has been discussed in the formalism with hidden 
or broken SU(4) symmetry. The superfield formalism based on the 
SU(3)/U(1)$\times$U(1) harmonics seems more flexible in the $N{=}4$ 
superspace.  A decomposition of the superfield constraints of the $N{=}4$ 
strength in terms of the 4-th spinor coordinate yields the equations of 
motion for the $N{=}3$ superfield strength $W^1$ and auxiliary superfield 
strengths $S_\da^1$ and $P^1$; however, one does  not succeed in 
constructing an appropriate superfield action in this setting.

As an illustration of flexibility of the $N{=}3$ superspace we have  
considered the new realization of the standard $N{=}4$ supersymmetry 
algebra on the infinite set of  $N{=}3$ superfields parametrized by the 
integer-number parameter $k$. This parameter plays a role of an additional 
discrete coordinate labeling different superfields of $N{=}4$ 
supermultiplets, and this degree of freedom is not connected directly with
any Grassmann or harmonic coordinates. It is not difficult to consider
the infinite-dimensional $N{=}4$ generalization of the $N{=}3$ gauge
harmonic-superspace action. This model has a simple geometric 
interpretation; however, the role of additional superfield degrees of 
freedom is still unclear and the consistency of this model at the quantum 
level is not verified. We hope to continue the investigation of possible 
realizations of the $N{=}4$ supersymmetry on a finite number of $N{=}3$ 
superfields by  analogy with the simplest example of such realization in 
subsection 4.2.
    
\vspace{3mm}

{\bf Acknowledgements}\\
The author is grateful to S. Ferrara for the  hospitality during the 
visit to Theory Division of CERN where  part of this work was made and to
E. Ivanov for discussions. The work was supported by grants INTAS  No 
00-00254, RFBR-DFG 02-02-04002, DFG No 436 RUS 113/669 and RFBR 
03-02-17440.

\def\theequation{A.\arabic{equation}}
\setcounter{equation}0
\section{Appendix. N=3 harmonic superspace}
In this appendix, one reviews basic formulae of the $D{=}4,~ N{=}3$ 
harmonic superspace \cite{GIKOS3,GIO,HS} using some results and the 
notation of Ref. \cite{IZ}. The basic coordinates of this superspace are 
the harmonics $u^I_i$ and $u_I^i$ parametrizing the 6-dimensional coset 
SU(3)/U(1)$\times$U(1)
\be
u^I_i u_J^i=\delta^I_J,\qquad
u^I_i u_I^j=\delta_i^j,\qquad
\varepsilon^{ijk}u^1_iu^2_ju^3_k=1
\ee
where $i,j,k=1,2,3$ are the indices of fundamental representations of the 
automorphism group SU(3), and $I,J=1,2,3$ are some combinations of 
U(1)$\times$U(1) charges in the condensed notation of Ref.\cite{IZ}. The 
SU(3)-invariant harmonic derivatives $\partial^I_J~ (I\neq J)$ act on 
these harmonics
\be
\partial^I_J u^K_i=\delta^K_Ju^I_i~,\q
\partial^I_J u_K^i=-\delta_K^Iu_J^i~.
\ee
Harmonic U(1) charges $h$ and $\tilde h$ arise in the commutators of 
the charged derivatives
\bea
&&[\pot,\pto]=h~,\q [\pth,\pht]=\tilde h~,\q[h,\pot]=2\pot~,\nn\\
&&[h,\pto]=-2\pto~,\q[\tilde h,\pth]=2\pth~,\q[\tilde h,\pht]=-2\pht~,
\lb{harmder}\\
&&[h,\pth]=-\pth~,\q[\tilde h,\pot]=-\pot~,\q [h,\poh]=\poh=[\tilde h,\poh]
~.\nn
\eea

The $N{=}3$ harmonics can be used to define the {\it analytic} coordinates
$\{x_\A^m,\theta_I^\alpha,\bar\theta^{I\dot\alpha},u\}$ in the harmonic 
superspace
\be
\begin{array}l
x_\A^m=x^m-
 i(\sigma^m)_\ada(\theta_1^\alpha\bar\theta^{1\dot\alpha}-
 \theta_3^\alpha\bar\theta^{3\dot\alpha}),\\
\theta_I^\alpha=\theta_i^\alpha  u^i_I, \qquad
\bar\theta^{I\dot\alpha}=\bar\theta^{\dot\alpha i}u^I_i
\end{array}
\ee
where $x^m, \theta^\alpha_i, \bar\theta^{i\da}$ are the standard
coordinates of the $D{=}4, N{=}3$ superspace and $\sigma^m$ are the Weyl 
matrices of the group SL(2,C).

We shall use the harmonic projections of the standard spinor derivatives 
$D^k_\alpha$ and $\bar D_{k\da}$
\bea
&&D^I_\alpha=u_k^ID^k_\alpha~,\q \bar D_{I\da}=u^k_I\bar D_{k\da}\nn\\
&&\{D^I_\alpha, D^K_\beta\}=0=\{\bar D_{I\da}, \bar D_{K\db}\}\\
&&\{D^K_\alpha, \bar D_{J\db}\}=2i\delta^K_J\padb~,\nn
\eea
where $\padb=(\sigma^m)_\adb \partial_m$. 

The special conjugation acts on the harmonics and harmonic derivatives
\bea
&&\widetilde{u^1_k}=u_3^k~,\q \widetilde{u^3_k}=u_1^k~,\q
\widetilde{u^2_k}=-u_2^k~,\nn\\
&&\widetilde{\pot}=-\pth~,\q \widetilde{\poh}=\poh~.
\eea
The  special conjugation of spinor derivatives can be chosen in the 
following form:
\be
\widetilde{D^1_\alpha f}=-\bar D_{3\da}\tilde f~,\q
\widetilde{D^2_\alpha f}=\bar D_{2\da}\tilde f~,
\ee
where $f$ and $\tilde f$ are mutually conjugated even harmonic 
superfields.

The $(4+4)$ Grassmann-analytic superfield  $\Lambda(\zeta,u)$ is defined 
in the  analytic $(4|4+4)$-dimensional superspace with the coordinates
\be
\zeta=(x^m_\A, \theta^\alpha_2, \theta^\alpha_3, \bar\theta^{1\da},
\bar\theta^{2\da})~.
\ee
The  superfield $\Lambda(\zeta,u)$ satisfies the conditions
\be
(D^1_\alpha , \bar D_{3\da})\Lambda=0~.
\ee
The harmonic derivatives in the analytic coordinates $D^I_K$ act 
covariantly on the spinor derivatives
\be
[D^I_K, D^J_\alpha]=\delta^J_KD^I_\alpha~,\q
[D^I_K,\bar D_{J\da}]=-\delta^I_J\bar D_{K\da}.
\ee

Note that the following harmonic and spinor derivatives preserve the 
Grassmann analyticity
\bea
&& D^1_2~,\q D^2_3~,\q D^1_3=[D^1_2, D^2_3]~,\q D^2_\alpha~,\q 
\bar D_{2\da}~.\lb{D2der}
\eea

The analytic  potentials $\Vot, \Vth$ and $\Voh$  of the $N{=}3$ 
Yang-Mills theory can be treated as gauge connections for the harmonic 
covariant derivatives
\bea
&&\nabla^I_K=D^I_K+V^I_K~,\q I<K~,\lb{lambda}\\
&&\delta_\lambda V^I_K=-D^I_K\lambda+[\lambda,V^I_K]~,\nn
\eea 
where $\lambda$ is an infinitesimal analytic gauge parameter. We use the 
following conditions of conjugation:
\be
\widetilde{V^1_2}=-V^2_3~,\q\widetilde{V^1_3}=V^1_3~,\q
\widetilde{\lambda}=-\lambda\lb{Vconj}
\ee 

The superfield action of the $N=3$ theory was  defined on these analytic 
potentials \cite{GIKOS3}
\bea
&&S(N{=}3)=-{1\over4}\int du\, d\zeta(^{33}_{11})\Tr \{V^1_2D^2_3V^1_3
+V^1_3D^1_2V^2_3+V^2_3D^1_3V^1_2-{1\over2}(V^1_3)^2\nn\\
&&+V^1_3[V^1_2,V^2_3]\}~.\lb{n3ym}
\eea
In this formula, the analytic integral measure is considered
\bea
&& d\zeta(^{33}_{11})=d^4x_\A d\theta(^{33}_{11})~,\q\int  
d\theta(^{33}_{11})(\theta_2)^2(\theta_3)^2(\bar\theta^1)^2
(\bar\theta^2)^2=1~.
\eea
The action yields equations of motion which have the form of 
zero-curvature conditions for the $N{=}3$ potentials
\bea
&&V^1_3=\Dot V_3^2-\Dth V^1_2+[V^1_2,V_3^2]~,\nn\\
&&F^{11}_{32}\equiv\Doh V^1_2-\Dot V^1_3
+[V^1_3,V^1_2]=0~.\lb{nazc}
\eea

The $N{=}3$  harmonic formalism allows us to introduce the off-shell  
superfield strengths \cite{IZ,NZ}
\bea
&&W^1={1\over4}(\bDh)^2\Vht(\Vth)~,\q \q\Dth\Vht-\Dht\Vth+[\Vth,\Vht]=0~,
\nn\\
&&\bar W_3=-{1\over4}(\Do)^2\Vto(\Vot)~,\q\Dot\Vto-\Dto\Vot+[\Vot,\Vto]=0
~,\lb{nonal}
\eea
which are defined via the nonanalytic connections $\Vto(\Vot)$ and
$\Vht(\Vth)$. The harmonic zero-curvature equations for the $N{=}3$ 
connections $V^I_K$ guarantee the harmonic analyticities of these 
superfields, for instance,
\be
\nabla^2_3W^1=\Dth W^1+[\Vth, W^1]=0~,\q\nabla^1_2W^1=\Dot W^1+[\Vot, 
W^1]=0~.\lb{nahan}
\ee
The first relation is valid off-shell, while the second one is equivalent 
to the equations of motion. 

It is not difficult to obtain  pseudodifferential solutions for the 
Abelian non-analytic connection
\be
\Vto={1\over2}(\Dto)^2\Vot-{1\over12}(\Dto)^3\Dot\Vot+\ldots
\ee
using the relation $[\Dot,\Dto]\Vot=2\Vot$. The superfield strength can be 
expressed via the potential
\bea
&&\bar W_3=-{1\over4}(\Do)^2\Vto=-{1\over4}(\Dt)^2P\Vot~,
\nn\\
&&P\equiv 1+\sum\limits^\infty_{k=1}(-1)^k{1\over k!(k+1)!}
(D^2_1)^k(D^1_2)^k~.\lb{pseudo}
\eea
An analogous formula for $W^1$ contains the conjugated operator $\tilde P$.

It is convenient to use the gauge conditions
\be
\Dot\Vot=0~,\qq \Dth\Vth=0\lb{gcond}
\ee
to simplify the pseudodifferential solutions of Eqs.\p{nonal}. In this 
case, one can construct a simple nonlinear expansion of the nonanalytic 
connection $\Vto$ in terms of the non-Abelian potential  
\bea
&&\Vto(\Vot)={1\over2}(\Dto)^2\Vot+{1\over4}[(\Dto)^2\Vot,\Dto\Vot]+
{1\over12}[\Dto\Vot,[\Dto\Vot,(\Dto)^2\Vot]]\nn\\
&&-{1\over24}[(\Dto)^2\Vot,[\Vot,
(\Dto)^2\Vot]]+\ldots\lb{naV}\\
&&\bar W_3(\Vot)=-{1\over4}(\Do)^2\Vto(\Vot)=-{1\over4}(\Dt)^2\Vot+\ldots
\lb{naW}
\eea
and analogous expansions for $\Vht(\Vth)$ and $W^1(\Vth)$.
 

\end{document}